\documentclass[onecolumn]{article}
\usepackage{authblk}
\usepackage{amsmath}
\usepackage{graphicx}
\usepackage{listings}
\usepackage{color}
\usepackage{subfig}

\usepackage{type1cm}
\usepackage{eso-pic}
\makeatletter
\AddToShipoutPicture{%
            \setlength{\@tempdimb}{.5\paperwidth}%
            \setlength{\@tempdimc}{.5\paperheight}%
            \setlength{\unitlength}{1pt}%
            \put(\strip@pt\@tempdimb,\strip@pt\@tempdimc){%
        \makebox(0,0){\rotatebox{45}{\textcolor[gray]{0.95}%
        {\fontsize{6cm}{6cm}\selectfont{DRAFT}}}}%
            }%
}
\makeatother

\definecolor{orange}{rgb}{1,0.6,0}
\definecolor{grey}{rgb}{0.4,0.4,0.4}
\title{Sinkless: A Preliminary Study of Stress Propagation in Group Project Social Networks
	using a Variant of the Abelian Sandpile Model}
\author{Bryan Knowles\thanks{bryan.knowles951@topper.wku.edu}\\
	Rong Yang\thanks{rong.yang@wku.edu}}
\affil{Computer Science\\Western Kentucky University\\Bowling Green, KY, USA}
\begin{document}
	\maketitle
	\begin{abstract}
		We perform social network analysis on 53 students split over three semesters and
		13 groups, using conventional measures like eigenvector centrality, betweeness
		centrality, and degree centrality, as well as defining a variant of the Abelian
		Sandpile Model (ASM) with the intention of modeling stress propagation in the
		college
		classroom. We correlate the results of these analyses with group project grades
		received; due to a small or poorly
		collected dataset, we are unable to conclude that any of these network measures
		relates to those grades. However, we are successful in using this dataset
		to define a discrete, recursive, and more generalized variant of the ASM.
		
	\end{abstract}
	\section{Introduction}\label{in}
		At Western Kentucky University, Technical Writing (ENG307) is a required course
		for students seeking degrees in Professional Writing, Computer Science,
		Engineering, Management, and so on. Dr. Angela Jones is one instructor
		of this course; in her sections of ENG307, she assigns a group project where
		groups of about four students construct some object with ``Legos" and compile
		an instruction manual for how to recreate that object. Front matter of this
		document would contain a parts list; body matter the instructions; and back matter
		a notes section. Groups are graded based on the best practices implemented in
		the manual, a presentation defending their choices, and orderly, timely submission
		of all project materials.
		
		Included in these project materials are project logs, in which students are
		asked to record what they did, when they did it, how long was spent doing it,
		and whether the task was completed individually or collaboratively. Dr. Jones
		uses these logs to determine whether a student did not sufficiently contribute
		to the groupwork.
		
		Interested in the effects of social network structure and location on college
		grades, we saw potential in the contents of these logs. As a preliminary work
		defining our models and methods, we received from Dr. Jones anonymous PDF scans
		of project logs for the past three semesters, fall 2012 to spring 2013. In all,
		there were 53 logs.
		
		To anonymize the logs, all names were replaced with unique identifiers (UID), such
		as ``F3." The first character of these UIDs represents the group to which
		the student belonged; the second character serves as an in-group identifier.
		
		In addition to the PDFs, we received a spreadsheet associating with each UID
		a semester, a grade, a gender, a major, and a year in college, such as
		freshman and senior.
		
		Groups were assigned by Dr. Jones randomly, adjustments being made if she suspected
		two or more students would be unable to work together.
		
			
	\section{Related Work}\label{rw}
		Sparrowe et al apply social network analysis (SNA) to 190 employees spanning five
		firms in \cite{sparrowe01}. They distinguish between positive and negative
		networks, or social networks where nodes are connected by positive and
		negative relations respectively. In particular, they analyze advice networks
		(positive) and hindrance networks (negative). Connections
		for these networks were determined by survey questions, such as
		``Do you go to [name] for help" or ``Does [name] make it difficult." Performance
		was measured by surveys answered by group leaders from each of 38 work groups.
		Sparrowe et al distinguish between individual-level performance and group-level
		performance, finding distinct correlations for each.
		
		They find that central members in advice networks are rated higher than others;
		conversely, those central in hindrance networks are rated lower; and
		paradoxically, highly central groups within a firm are rated lower. Conclusions
		were based on in-degree centrality, network density, and network centralization
		measures.
		
		Preparing to perform a similar analysis for our dataset on group projects,
		we come
		across the abelian sandpile model (ASM) in the work of D'Agostino et al
		\cite{dagostino12}, where it has been adapted for research on financial
		distress propagation.
		
		Bak et al define the ASM in \cite{bak87}. Given an infinite
		2-dimensional grid, assume ``grains of sand" have been dropped thereonto.
		The amount of sand at a location $(x, y)$ being denoted by $z(x, y)$ and
		given a critical value $K$, we update each cell of the grid simultaneously
		according to equation \ref{eq:original}. At least one node must be reachable
		by a path extending from each other node and have $K = \infty$; such nodes are
		said to be ``sinks" because sand ``drains out" through them.
		
		\begin{equation}\label{eq:original}
			\begin{split}
				z(x, y) \rightarrow z(x, y) - 4, \\
				z(x\pm1, y) \rightarrow z(x\pm1, y) + 1, \\
				z(x, y\pm1) \rightarrow z(x, y\pm1) + 1, \\
				\text{if } z(x, y) > K.
			\end{split}
		\end{equation}
		
		Performing a simulation on this model involves initializing the grid
		with a random distribution of sand, allowing the sand to propagate according
		to equation \ref{eq:original} until the cascade halts, and probing the
		system by sequentially dropping additional grains of sand.
		
		To apply this model to financial distress, D'Agostino et al adapt it
		to operate on networks of any structure: when a node has received an amount
		of sand exceeding its threshold, it ``topples," donating one of its grains
		to each of its neighbors. Next, they let sinks represent banks under the
		coverage of a central bank; that is, these banks are ``immunized" against
		bankruptcy.
		
		They construct scale-free networks, or informally those produced by a
		``rich get richer" effect. Analyzing the frequencies of cascades of various
		sizes, they conclude that networks whose degree-degree correlation is
		positive are more likely to experience a pandemic, yet the speed of pandemic
		diffusion is decreased.
		
		Others have made similar adaptations of the ASM, including applications
		in modeling blackouts \cite{carreras01}, data-packet transportation
		\cite{lee05}, rivers \cite{prigozhin94}, and superconductors \cite{ginzburg00}.
		
		In this paper, in section \ref{ssm} particularly, we define a variant of the
		ASM that achieves self-organized critical behaviors without the
		requirement for sinks.
	
	\section{Experimental Models}\label{em}
		\subsection{Friend Approximation Network}
			Because the project logs were hand-written by students and
			inconsistently filled out, we were unable to use these to define
			a social network. Turning to the spreadsheet filled out by Dr. Jones,
			we assume for the sake of this preliminary work that two students
			are friends if they either (1) are in the same group or (2) share
			the same college year and major; students may only
			be connected if they are in the same Technical Writing class.
			These assumptions are based on intuition that
			students who share majors and years are more likely to have other
			classes together and that
			students who communicate often are more likely to be friends.
			
			We define our social network, denoted by an
			adjaceny matrix $G$, in equation \ref{eq:G}.
			
			\begin{equation}\label{eq:G}
				G_{ij} = \begin{cases}
					1, & \text{GROUP(i)=GROUP(j)}\\
					1, & \text{MAJOR(i)=MAJOR(j)} \wedge \text{YEAR(i)=YEAR(j)} \\
					0, & \text{Otherwise}
				\end{cases}
			\end{equation}
			
			We refer to $G$ as a ``friend approximation network" (FAN) and
			say that two students, $i$ and $j$ are ``approximately friends"
			iff $G_{ij}=1$; approximate friendship is undirected, reflected
			here by $G_{ij} = G_{ji}$.
			
			\begin{figure}
				\centering
				\includegraphics[width=0.4\textwidth]{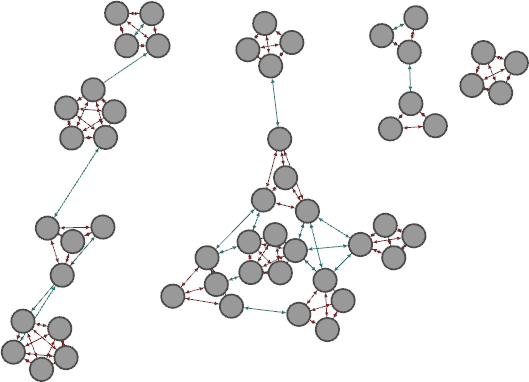}
				\caption{The FAN, containing 53 nodes and 202 edges.
				Groups form complete subgraphs and are loosely connected
				to other groups. The leftmost subgraph represents the
				FAN for fall 2012, the center for spring 2012, and the
				rightmost two for spring 2013.}
				\label{fig:FAN}
			\end{figure}
			
		\subsection{Sinkless Sandpile Model}\label{ssm}
			Because students cannot support an unlimited amount of stress
			and because we wish not to introduce additional nodes
			that can, we remove sinks from our sandpile model.
			So to prevent infinite loops from occuring, we allow nodes
			to ``blow away" a fraction of a grain of sand each time they
			would topple. Before a node $i$ topples, the sand at $i$ is
			decreased by some parameter $0 < g < 1$. Interestingly, if $g=1$,
			our model is equivalent to an ASM with a sink connected directly
			to each node.
			
			We denote the network-level carrying capacity (CC), or maximum
			amount of sand holdable before a topple must occur, as $K$. We
			denote the CC for node $i$ as $k_i$; that is, $k$ is a vector
			of node-level CCs.
			
			To determine $K$ and $k$, we can either assign all $k_i$
			values and sum them to receive $K$ or assign $K$ and
			distribute that sum over all $k_i$ values. We find that the second
			option allows easier comparison of results from networks of
			different size and structure.
			
			We define $k_i$ at first based on a share of $K$ proportionate to node $i$'s
			degree. Equation \ref{eq:K} shows this basic definition, where $DEG(i)$ is
			the number of neighbors of $i$.
			
			To generalize further, we introduce a parameter $P$ into that equation,
			producing equation \ref{eq:KP}. When $P=1$, the CC distribution is relative
			to the degree distribution of the network; when $P=0$, CCs are equally
			distributed; when $P>1$ or $P<0$, the CC is disproportionately
			distributed relative to degree distribution, where $P>1$ intensifies
			the relationship and $P<0$ inverses it.
			
			\begin{subequations}
				\begin{equation}\label{eq:K}
					k_i = K \frac{DEG(i)}{\sum_j DEG(j)}
				\end{equation}
				\begin{equation}\label{eq:KP}
					k_i = K \frac{DEG(i)^P}{\sum_j \left(DEG(j)^P\right)}
				\end{equation}
				\begin{equation}\label{eq:minK}
					K \ge (min(DEG) + g) * \sum_j(DEG(j)^P) \div \begin{cases}
						min(DEG) & P \ge 0 \\
						max(DEG) & P < 0
					\end{cases}
				\end{equation}
			\end{subequations}
			
			\begin{figure}
				\centering
				\subfloat{%
					\includegraphics[width=0.5\textwidth]%
					{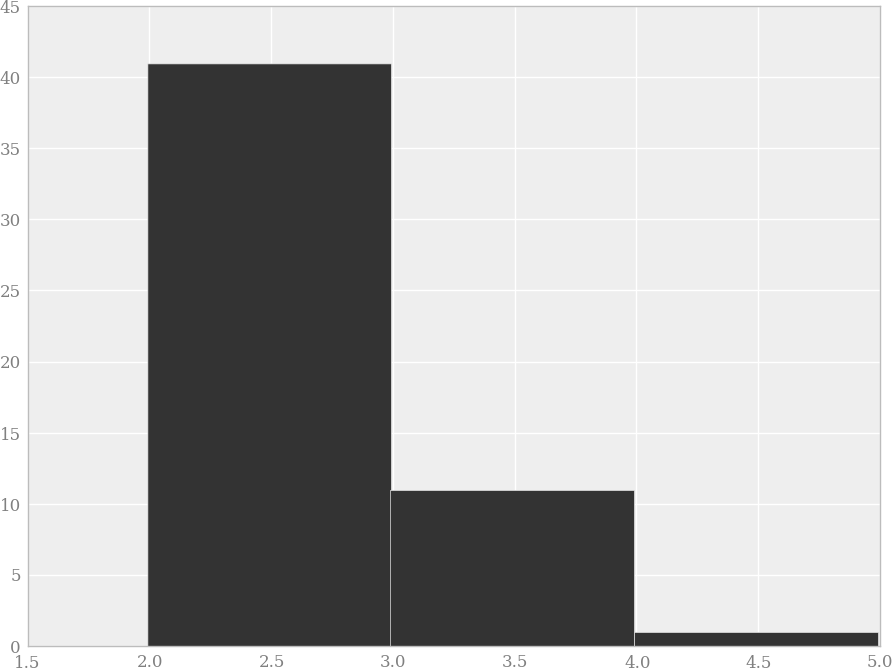}%
				}
				\subfloat{%
					\includegraphics[width=0.5\textwidth]%
					{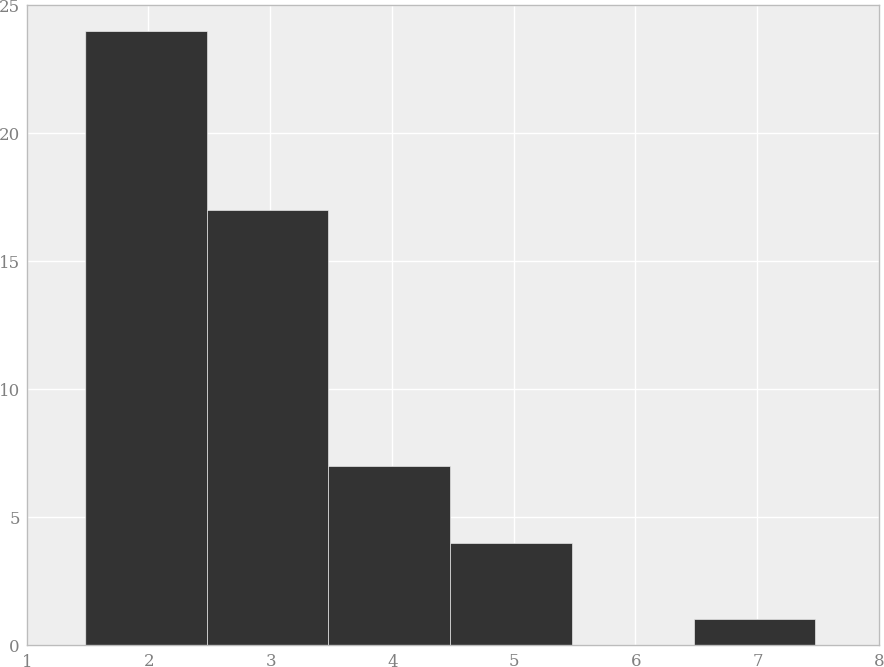}%
				}
				\caption{CC distribution for the friend approximation network
				when $P=1$ (left) and $P=2$ (right)}
				\label{fig:CCdist}
			\end{figure}
			
			We require that $k_i \ge DEG(i)+g$ for all $i$ in order to prevent a
			topple from reducing the amount of sand at any node to a negative
			value. The minimum acceptable $K$ follows from equation \ref{eq:minK}.
			
			We denote the number of grains to drop in the simulation as $X$.
			In order to provide enough time for the model to reach self-organized
			criticality \cite{bak87}, we require $X \gg K$.
			
			We denote the amount of sand at node $i$ and time $t$ as $S(i,t)$.
			We denote the neighbors of $i$ as $N(i)$. We denote the nodes toppling
			at time $t$ as $T(t)$ and define it in equation \ref{eq:T}. We
			denote the toppling neighbors of $i$ at time $t$ as $TN(i, t)$
			and define it in equation \ref{eq:TN}.

			\begin{subequations}			
				\begin{equation}\label{eq:T}
					T(t) = \left\{i|S(i,t) > k_i\right\}
				\end{equation}
				\begin{equation}\label{eq:TN}
					TN(i,t) = N(i) \cap T(t)
				\end{equation}
			\end{subequations}
			
			We denote a random sequence of integers inclusively between $1$ and the
			number of nodes as $RND$. So that grains are not dropped during a
			cascade, we define $ZRND$ in equation \ref{eq:rnd2}.
			
			Our model may then be defined by the recurrence relation in equation
			\ref{eq:model}.

			\begin{subequations}			
				\begin{equation}\label{eq:rnd2}
					ZRND_t = \begin{cases}
						RND_t & T(t) = \emptyset \\
						0 & \text{Otherwise}
					\end{cases}
				\end{equation}
				\begin{equation}\label{eq:model}
					\begin{split}
						S(i,t+1) = S(i,t) + |TN(i,t)| \\
						+ \begin{cases}
							1 & ZRND_t = i \\
							0 & \text{Otherwise}
						\end{cases} - \begin{cases}
							DEG(i)+g & i \in T(t) \\
							0 & \text{Otherwise}
						\end{cases}
					\end{split}
				\end{equation}
			\end{subequations}
			
			We ``halt" $S(i,t)$ at the earliest $t$ where the first $t$ elements of
			$ZRND$ contain $X$ nonzero values.
			
			Summarizing informally, we drop $X$ grains of sand sequentially onto the
			network.
			When a node has received an amount of sand greater than its CC, it
			topples. When a node topples, its amount of sand is first decreased
			by $g$ and then, for each neighbor of that node, a single grain is taken
			from it and given to that neighbor. If multiple nodes would topple
			at once, their amounts of sand are updated synchronously.
			
			A python implementation of this process is given in Appendix A.
		
	\section{Experimental Results}\label{er}
		We performed 1,000 simulations each for 20 different configurations of
		the above model, with $X=2500$, $g=0.1$, $K=880$, $P\in[-2..2]$, and $G$
		representing any of the three semesters or all three combined. In all,
		50,000,000 virtual grains of sand were dropped. We recorded the number
		of topples that occured for each node during each simulation and the
		network-total number of topples (NTNT) that occured after each grain of sand
		during each simulation.
		
		Nearly identical patterns emerged in all 20,000 simulations; in particular,
		the average rate of growth of NTNT for each configuration
		of parameters converged to the same linear trend after about 2,300 grains
		were dropped and propagated. The observed NTNT for any single simulation
		behaved unpredictibly, albeit following this linear trend. Before
		the critical point of 2,300 grains, each configuration behaved uniquely;
		each simulation for a given configuration tended towards the same behavior
		regardless of how many grains had been dropped.
		
\begin{figure}
			\centering
			\subfloat{%
				\includegraphics[width=0.3\textwidth]%
				{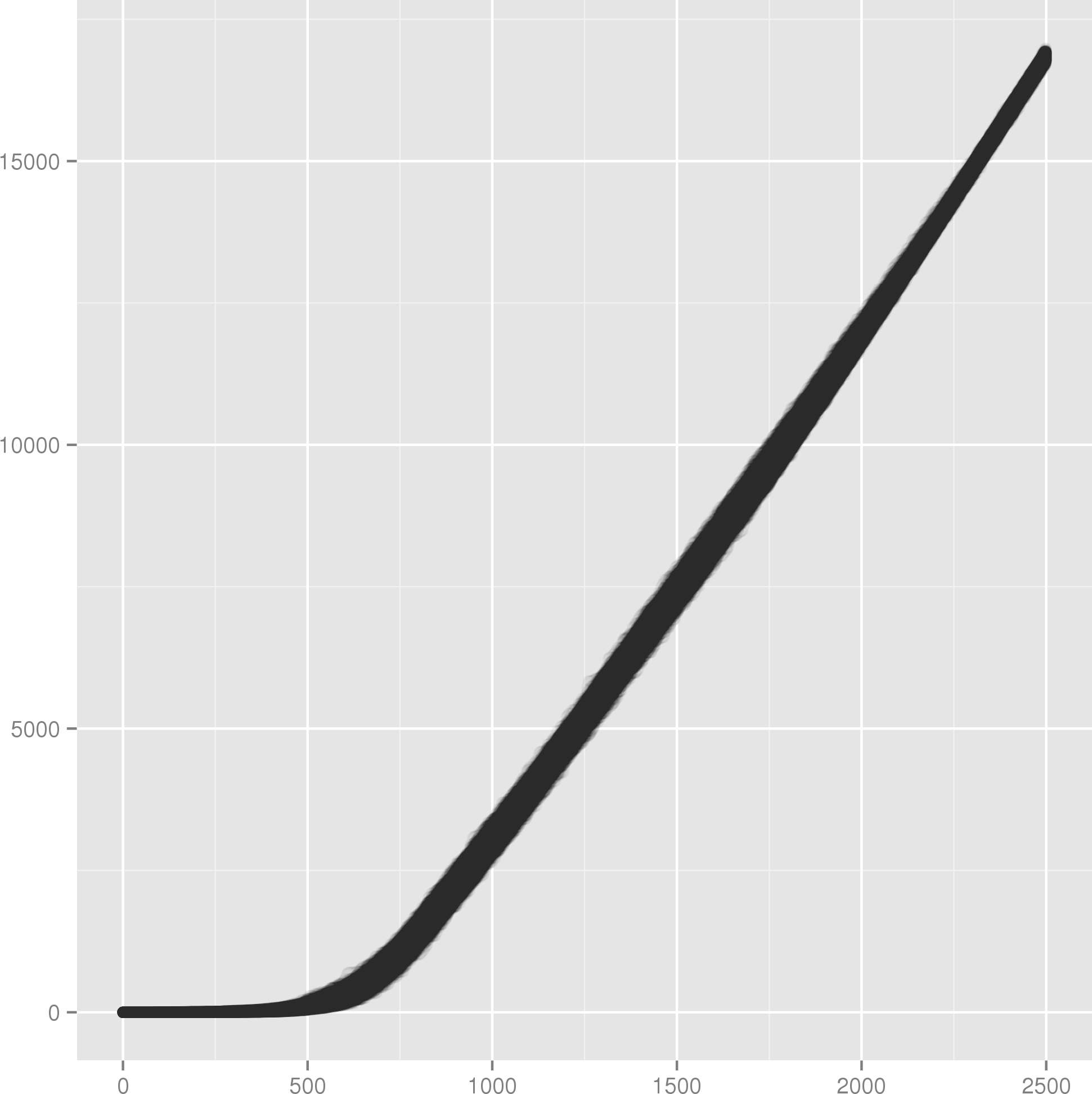}%
			}
			\subfloat{%
				\includegraphics[width=0.3\textwidth]%
				{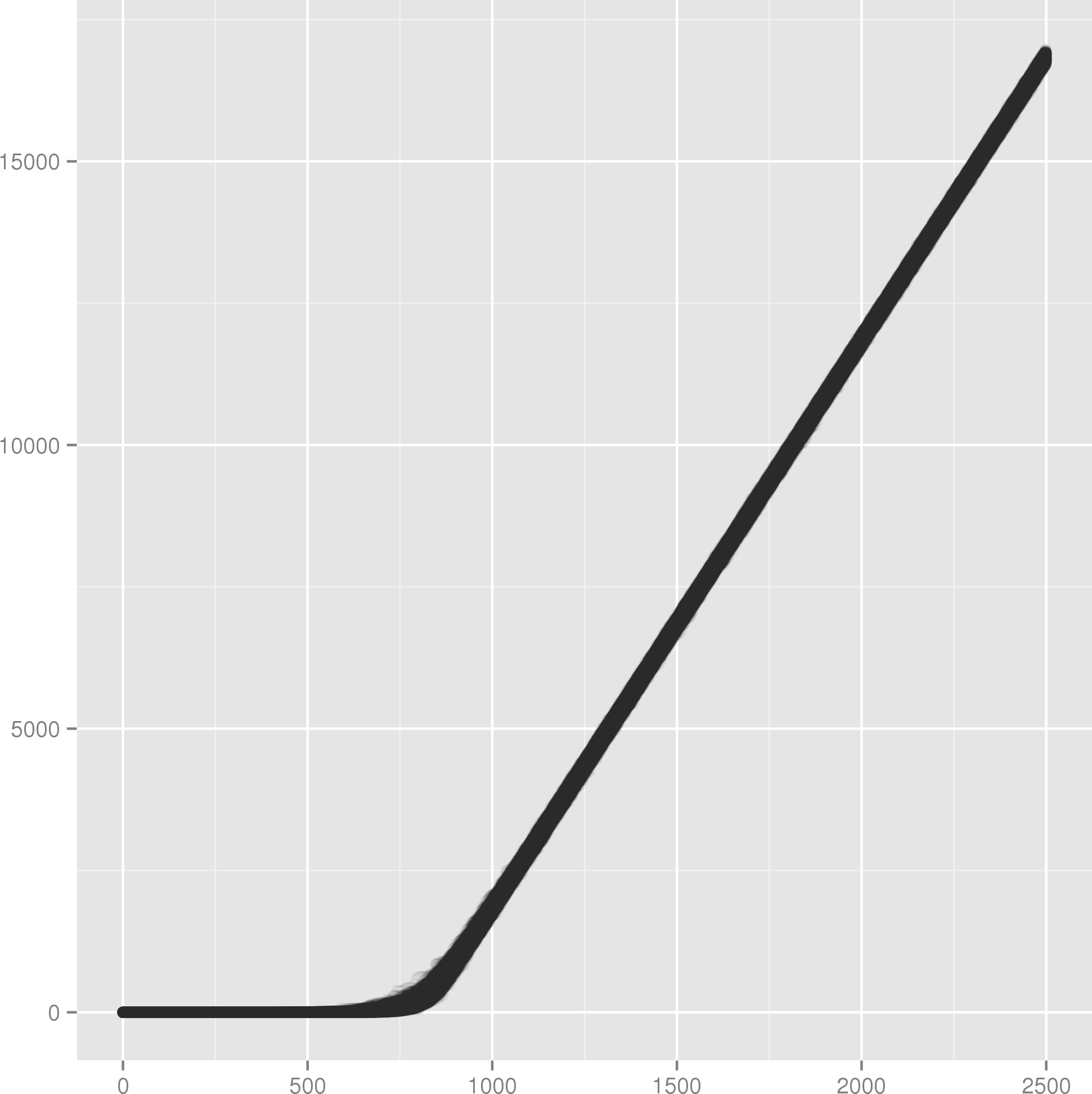}%
			}
			\subfloat{%
				\includegraphics[width=0.3\textwidth]%
				{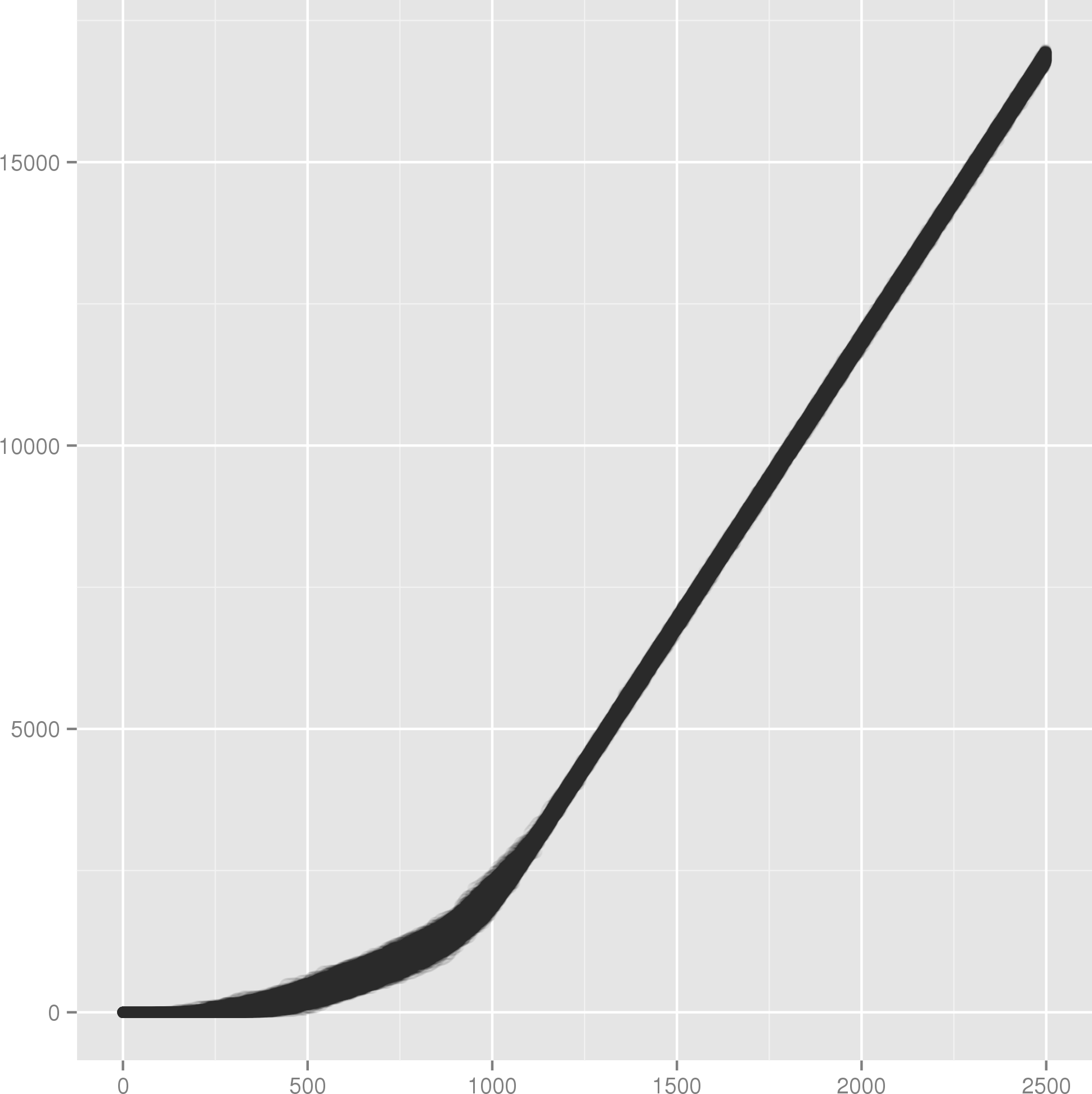}%
			}
			\caption{NTNT growth for full dataset where $P=-2$ (left),
			$P=0$ (middle), and $P=2$ (right). Points are plotted for all
			1,000 simulations at 1\% opacity each.}
			\label{fig:NTNT}
		\end{figure}

		Figure \ref{fig:NTNT} provides a few examples.
		
		Plotting topples occured against grades received, we notice a few weak
		trends. When $P=0$, students that recieved As, Bs, and Cs all averaged
		close to the same number of topples, regardless of other parameters.
		
		When $P=1$ in the fall 2012 network, Bs experienced the most topples and
		Cs were a close third; in the spring 2012 network, Bs still experienced
		the most, but Cs were in second and the deviation between the three letter
		grade groups was small; in the spring 2013 network, Bs experienced far
		fewer topples than either As or Cs, who were almost identical; and in the
		full network, results were similar to that of the spring 2012 network
		with less deviation between the three letter grade groups. We note that
		the spring 2012 network was the largest and made up roughly 47\% of the
		full network.
		
		When $P<0$, the order of the results were the reverse of those of when
		$P>0$. When $|P|$ increases, the deviation between the number of topples
		received increases between the three letter grade groups.
		
\begin{figure}
			\centering
			\subfloat{%
				\includegraphics[width=0.5\textwidth]%
				{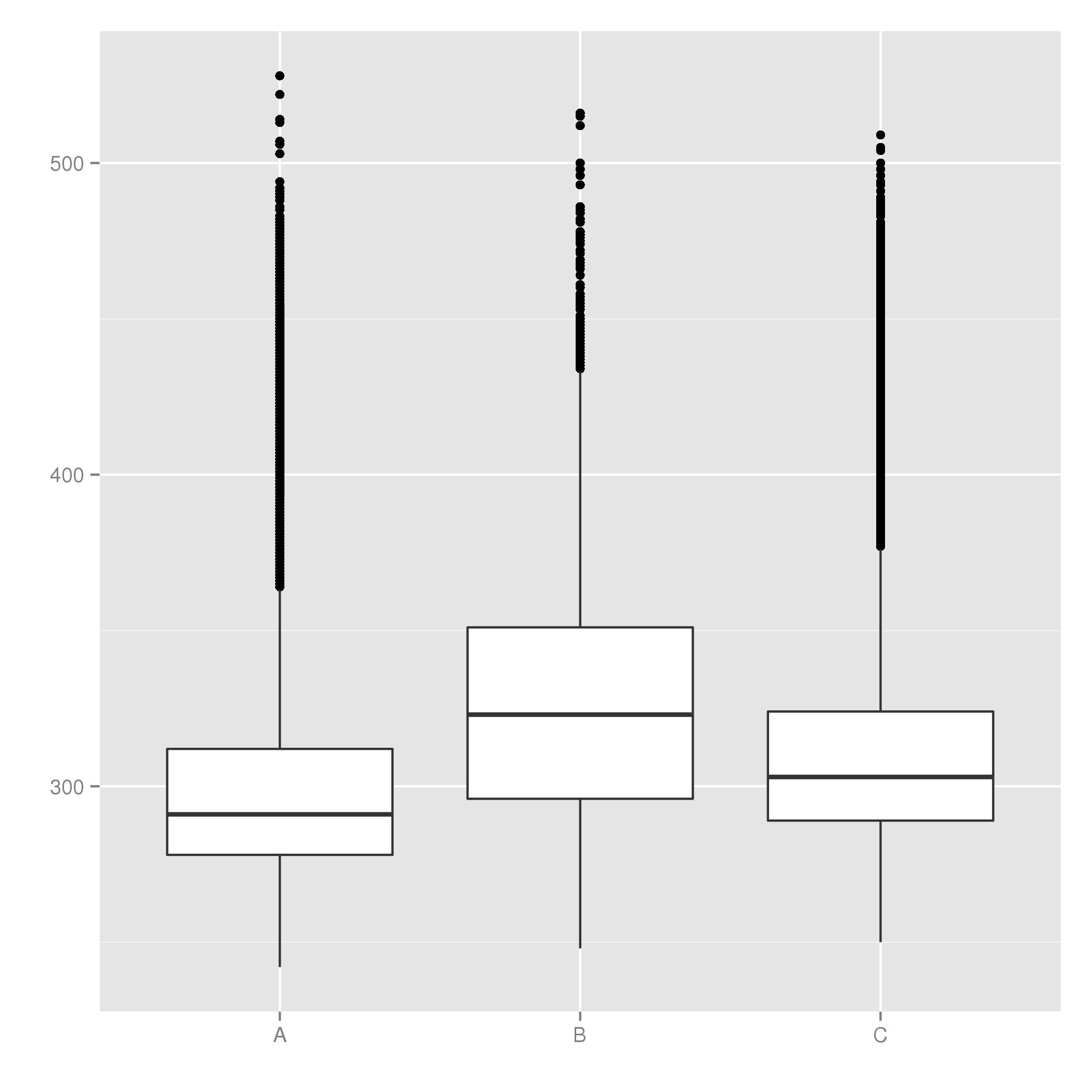}%
			}
			\subfloat{%
				\includegraphics[width=0.5\textwidth]%
				{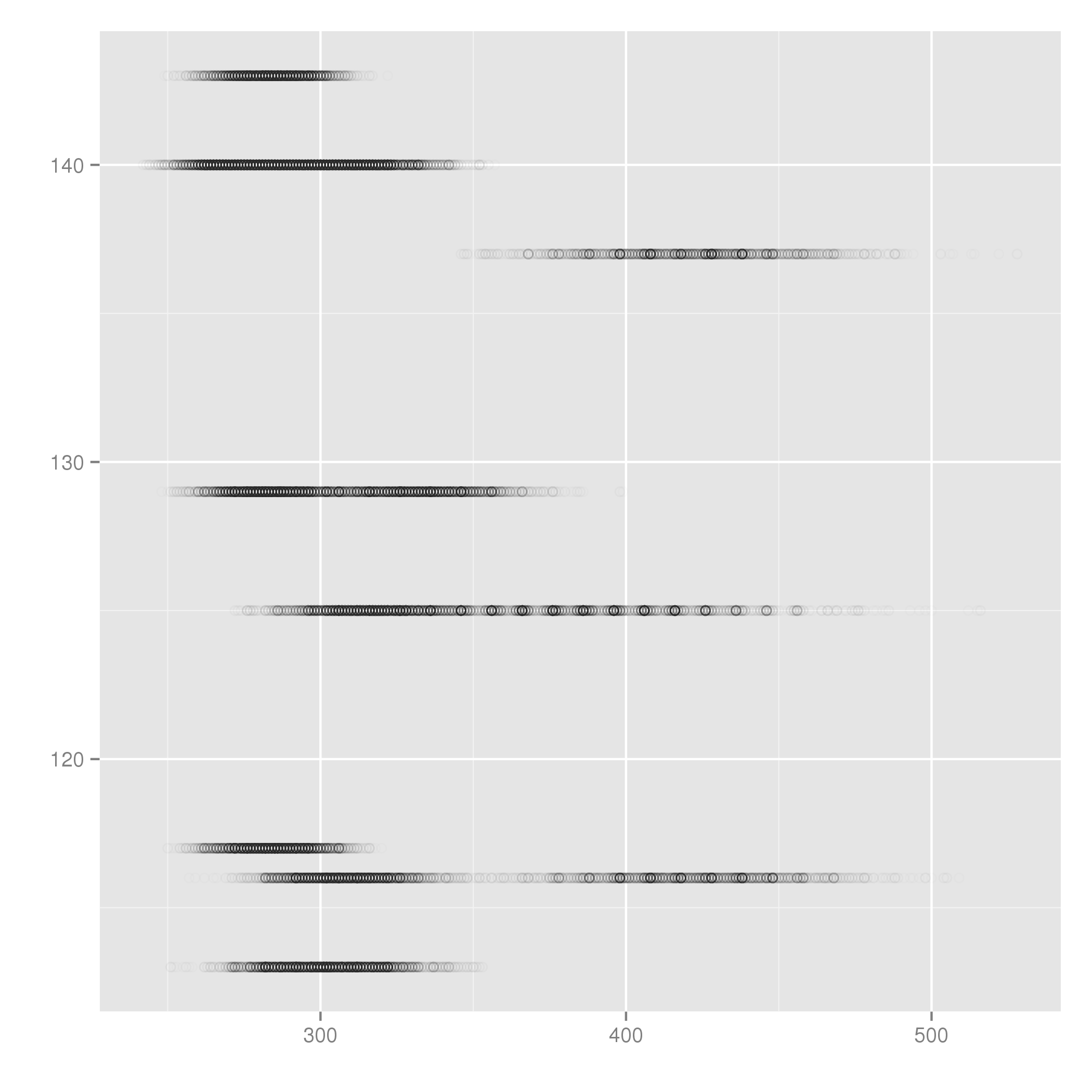}%
			}
			\caption{Plots for topples occured against grades received for the
			full network, $P=2$, and all 1,000 simulations. Points on the right are
			plotted at 1\% opacity each.}
			\label{fig:boxes}
		\end{figure}

		Figure \ref{fig:boxes} shows the results for one configuration. Note the
		grouping of grades into three distinct groups, corresponding to As, Bs,
		and Cs. This is an effect of Dr. Jones's grading procedure on projects:
		roughly, she first determines whether the submission is A-work, B-work, or
		C-work; next, she assigns a numerical grade producing that letter
		grade; finally, she adjusts based on details like spelling and whitespace
		management.
		
	\section{Comparisons}\label{co}
		The results from our simulations were weak, so to compare out method's
		performance with that of common SNA measures, we
		calcualte Pearson correlation cofficients of grades vs. measure values
		for seven metrics in additional to our own.
		
		We, like \cite{sparrowe01}, divide these into two levels: member- and
		group-level.
		Member-level measures included topples occured and eigenvector, degree,
		and betweeness centrality, as defined in \cite{diestel05,freeman77,borgatti05}.
		Group-level measures included average intergrade, average year,
		gender ratio, and size. Because of the relatively fixed
		group size of about four members, the last two of these were
		uninteresting.
		
		Although the project logs were unusable in constructing a social
		network, they contained information providing additional insights.
		Students were asked to grade one another in their groups. To avoid
		confusion with individual or group grades, we refer to these as
		``intergrades." The average intergrade for a group then is the mean
		of the intergrades assigned to its members by its members.
		
		The average year of a group is simply the mean of values assigned to
		its members based on college year: freshmen recieved $1$s, sophomores
		received $2$s, and so on.
		
		All eight of these measures produced correlations weaker
		than $\pm.3776$. They are summarized in table \ref{table:rho}.
		
		\begin{table}
			\centering
			\begin{tabular}{lrlr}
				\hline
				\hline
				\multicolumn{2}{c}{Member-level} &
				\multicolumn{2}{c}{Group-level} \\
				\hline
				Metric & $\rho$ & Metric & $\rho$ \\
				\hline
				Topples & $-.1930$ & AvgIntergrade & $+.3073$ \\
				Eigenvector & $+.3776$ & AvgYear & $+.2278$ \\
				Degree & $+.2187$ & Gender & $-.1712$ \\
				Betweeness & $+.1926$ & Size & $+.1392$ \\
				\hline
			\end{tabular}
			\caption{Pearson correlation coefficients for each of eight metrics
			against grades received. Full network data was used.}
			\label{table:rho}
		\end{table}
		
	\section{Conclusions}\label{cn}
		We are unable to report a correlation between stress propagation as modeled
		by our variant of the ASM; further, we are unable
		to report correlations between SNA metrics that have
		received considerable and growing attention in the field \cite{borgatti05}.
		
		Because these SNA measures have been found to correlate
		strongly with performance in other data, both positively and negatively
		depending on the given network \cite{sparrowe01}, we believe that our failure
		in finding a
		correlation among these methods suggests either the dataset is too small,
		the collection methods were flawed, our friend approximation
		network was based on unsound assumptions, or some combination thereof.
		
		We were aware of these issues early on; however, obtaining
		rigorous datasets on student grades is nontrivial given concerns of privacy
		and the 1974 act, FERPA.		
		Instead for this preliminary work, we have taken our time to define and
		explore a novel adaptation of a ``sinkless" sandpile model. We define it
		using recurrence relations and discrete timesteps, allowing it to be easily
		further adapted for deterministic applications. Additionally, the ASM is
		reducible to our model by assigning $g=0$ and at least one $k_i = \infty$.
		Alternate strategies of determining $k$ may be implemented by making changes
		to equation \ref{eq:KP}; doing so requires care to ensure $k_i \ge DEG(i)+g$
		for all nodes $i$.
		
		Beyond our interests in student grades, then, we hope to provide a more general
		model of self-organized criticality to the field \cite{bak87}.

	\section{Special Thanks}\label{st}
		We would like to thank Western Kentucky University
		for supporting this work through a Faculty/Undergraduate
		Student Engagement (FUSE) grant; Dr. Jones
		for her hard work anonymizing and digitizing student logs, as well
		as preparing the spreadsheet that made this project possible; and
		Benjamin Thornberry for his always helpful assistance and feedback.
	
	\bibliographystyle{plain}
	\bibliography{references.bib}

\begin{thebibliography}{10}

\bibitem{bak87}
Per Bak, Chao Tang, and Kurt Wiesenfeld.
\newblock Self-organized criticality: An explanation of 1/f noise.
\newblock {\em Phys. Rev. Lett.}, 59:381--384, 1987.

\bibitem{borgatti05}
Stephen Borgatti.
\newblock Centrality and network flow.
\newblock {\em Social Networks}, 27:55--71, 2005.

\bibitem{carreras01}
B.~Carreras, V.~Lynch, M.~Sachtjen, I.~Dobson, and D.~Newman.
\newblock Modeling blackout dynamics in power transmission networks with simple
  structure.
\newblock In {\em Hawaii International Conference on System Sciences}. IEEE,
  2001.

\bibitem{dagostino12}
G.~D'Agostino, A.~Scala, V.~Zlatic, and G.~Caldarelli.
\newblock Robustness and assortativity for diffusion-like processes in
  scale-free networks.
\newblock {\em EPL}, 97, 2012.

\bibitem{diestel05}
Reinhard Diestel.
\newblock {\em Graph Theory}.
\newblock Springer-Verlag, 3rd edition, 2005.

\bibitem{freeman77}
Linton Freeman.
\newblock A set of measures of centrality based upon betweenness.
\newblock {\em Sociometry}, 40:35--41, 1977.

\bibitem{ginzburg00}
S.~Ginzburg and N.~Savitskaya.
\newblock Self-organization of the critical state in granular superconductors.
\newblock {\em Journal of Experimental and Theoretical Physics}, 90:202--216,
  2000.

\bibitem{lee05}
E.~Lee, K.~Goh, B.~Kahng, and D.~Kim.
\newblock Robustness of the avalanche dynamics in data-packet transport on
  scale-free networks.
\newblock {\em Phys. Rev. E}, 71, 2005.

\bibitem{prigozhin94}
L.~Prigozhin.
\newblock Sandpiles and river networks: Extended systems with nonlocal
  interactions.
\newblock {\em Phys. Rev. E}, 49, 1994.

\bibitem{sparrowe01}
Raymond Sparrowe, Robert Liden, Sandy Wayne, and Maria Kraimer.
\newblock Social networks and the performance of individuals and groups.
\newblock {\em Academy of Management}, 44:316--325, 2001.

\end{thebibliography}

	\clearpage
	\section*{Appendix A: Python Implementation}
		\begin{lstlisting}
def simulation(G, X, g, k):
	# initialization
	N = len(G)
	S = [0.0 for i in range(N)]
	T = [0 for i in range(N)]
	deg = [sum(G[i]) for i in range(N)]
	# drop X grains sequentially
	for x in range(X):
		i = randint(0, N-1)
		S[i] += 1
		# if i topples, assume a cascade
		if S[i] > k[i]:
			# until cascade stops...
			while True:
				# find topplers
				A = [j for j in range(N) if S[j] > k[j]]
				if len(A) == 0:
					break
				else:
					# find how to move sand
					B = []
					for j in A:
						T[j] += 1 # tally topple
						S[j] -= g # blow away
						S[j] -= deg[j]
						C = [m for m in range(N) if G[j][m]==1]
						B.extend(C)
					# move sand
					for b in B:
						S[b] += 1
	return T
			\end{lstlisting}
	
\end{document}